\documentclass[lettersize,journal]{IEEEtran}
\usepackage{hyperref}
\usepackage{amsmath,amsfonts}
\usepackage{amsmath}
\usepackage[sort]{cite}
\usepackage{amsfonts}
\usepackage{array}
\usepackage{color, amssymb}
\usepackage[caption=false,font=normalsize,labelfont=sf,textfont=sf]{subfig}
\usepackage{textcomp}
\usepackage{mathrsfs}
\usepackage{verbatim}
\usepackage{graphicx}
\def\BibTeX{{\rm B\kern-.05em{\sc i\kern-.025em b}\kern-.08em
    T\kern-.1667em\lower.7ex\hbox{E}\kern-.125emX}}
\usepackage{balance}
\usepackage{algorithm}
\usepackage{algorithmic}

\begin{document}
\title{\huge Active Cell Balancing for Extended Operational Time of Lithium-Ion Battery Systems in Energy Storage Applications}

\author{Yiming~Xu,~\IEEEmembership{Student Member,~IEEE},
\and Xiaohua~Ge,~\IEEEmembership{Senior Member,~IEEE},
\and Ruohan~Guo,~\IEEEmembership{Student Member,~IEEE},\\
and \and Weixiang~Shen,~\IEEEmembership{Senior Member,~IEEE}

\thanks{This work was partially supported by the Australian government research
training program scholarship offered to the first author of this study
.}

\thanks{Y. Xu, X. Ge, R. Guo, W. Shen are with the School of Science, Computing and Engineering Technologies, Swinburne University of Technology, Melbourne, VIC 3122, Australia (email: yimingxu@swin.edu.au; xge@swin.edu.au; rguo@swin.edu.au; wshen@swin.edu.au).

Preprint submitted to IEEE Transactions on Transportation Electrification on 02-May-2024.}
}

\maketitle

\begin{abstract}
Cell inconsistency within a lithium-ion battery system poses a significant challenge in maximizing the system operational time. This study presents an optimization-driven active balancing method to minimize the effects of cell inconsistency on the system operational time while simultaneously satisfying the system output power demand and prolonging the system operational time in energy storage applications. The proposed method utilizes a fractional order model to forecast the terminal voltage dynamics of each cell within a battery system, enhanced with a particle-swarm-optimisation-genetic algorithm for precise parameter identification. It is implemented under two distinct cell-level balancing topologies: independent cell balancing and differential cell balancing. Subsequently, the current distribution for each topology is determined by resolving two optimization control problems constrained by the battery's operational specifications and power demands. The effectiveness of the proposed method is validated by extensive experiments based on the two balancing topologies. The results demonstrate that the proposed method increases the operational time by 3.2$\%$.
\end{abstract}


\section{Introduction}

\IEEEPARstart{L}{ithium-ion} batteries (LIBs) have emerged as a desired power source for electrified transportation and energy storage systems (ESS), owing to their high energy and power density, low self-discharge rate and long lifespan \cite{JTCCESM2023,KSKMTIV2020, QYCWET2023, QYCXJES2023}.
To satisfy voltage level demands, LIBs are commonly configured in series to compose a battery system.
Nevertheless, inherent cell inconsistency stemming from manufacturing lead to the imbalances in state of charge (SOC) and terminal among battery cells \cite{MWMAJPS2023, KZLJASME2023, GWYSJES2023}.
Such imbalances can be detrimental, as the overall battery system capacity is constrained by the weakest cell \cite{YYJHTIE2024, KMRKTEC2023, YZSE2020}. Consequently, rectifying these imbalances becomes imperative for prolonging the operational time of a battery system.

To counteract these imbalances in cell voltages or SOCs, battery cell balancing is performed by applying variable currents to cells within a battery system, which can be roughly classified into two types: passive (dissipative) balancing and active (nondissipative) balancing \cite{ATEnergy2023}. The former uses resistors to dissipate energy from the cells that have a higher voltage or a higher SOC during charging\cite{YWXHJPS2023}.
The latter uses power electronics circuit to transfer energy from stronger cells to weaker cells \cite{YWPLJES2023}. Although passive cell balancing is often implemented in a battery system due to cost efficiency, it can be only conducted in battery charging. On the other hand, active cell balancing allows cells to be balanced at the end of a driving cycle and can achieve balancing in battery discharging which can considerably prolong the operational time of a battery system \cite{HWTPEL2023}.

Great efforts have been devoted to the equalization topology \cite{NGXHTPE2021}. However, the emphasis of this study will be placed on improving system-level battery performance via advanced control strategies. Many active equalization algorithms have been studied in the literature, such as rule-based \cite{ASSCJES2021}, fuzzy logic-based \cite{MEMRJES2019}, multi-agent consistency control \cite{QOJCTSE2018, NYLHTTE2023}, model predictive control (MPC) methods \cite{LMMPTIE2017, FHCST2022}, etc.
For instance, Ouyang et al. \cite{QOTII2020} introduced a hierarchical optimal equalization control algorithm focused on reducing equalization time and suppressing cell temperature. Ma et al. \cite{YMPDTIE2018}
applied a fuzzy logic control to implement the SOC-based equalization scheme through a two-stage bidirectional equalization circuit for battery inconsistency improvement and equalization efficiency optimization.
Ouyang et al. \cite{QOTTE2022} proposed a two-layer MPC strategy that independently controls the module-level and the cell-to-module-to-cell equalization currents, demonstrating the advantages of a hierarchical structure in minimizing computational load and maintaining cell current limitation.
Chen et al. \cite{JCABTAS2023} developed three MPC-based strategies to follow a uniform trajectory, maximizing the lowest cell voltage and minimizing the cell discrepancies.
Fan et al. \cite{TFENergy279} developed an MPC algorithm incorporating a fast-solving strategy to determine the equalization current for battery systems.
Ouyang et al. \cite{QOTIE8} presented a quasi-sliding-model-based control strategy to regulate the equalization current via a bidirectional modified Cuk converter.
Dong et al. \cite{GDTII2021} formulated the cell equalization problem as a path-searching problem, solved by the A-star algorithm to find the shortest path, corresponding to the most efficient SOC equalization.

Unlike the active battery cell balancing methods mentioned-above that primarily focus on voltage-based and SOC-based equalization, this study revisits the cell balancing problem in a battery system but presents an active optimization-based balancing method to extend battery system operational time, taking into account power demands and other critical factors. The main contributions of this paper are summarized as follows.

(1) We establish a fractional-order battery system model that incorporates an active balancing mechanism, using particle swarm optimization-genetic algorithm (PSO-GA) parameter identification. This modeling approach can effectively characterize the dynamics of the battery system, thus laying a robust base for accurate state estimation.

(2) We develop an efficient optimization-based algorithm for active cell balancing to extend the operational time of the battery system while adhering to voltage bounds, current limitations and power demands. The determination of the optimal current distribution is achieved through the solution of a multi-constraint optimization problem.

(3) We validate the effectiveness of the proposed balancing algorithm via the experiments under two battery systems using the current profiles based on the Urban Dynamometer Driving Schedule (UDDS). The enhanced performance has been evidenced by a comparative analysis of cell terminal voltages within a battery system and operational time before and after balancing.

The rest of this article is organized as follows. The battery model is described in Section II. The control strategy is presented in Section III. The balancing algorithm is introduced in Section IV. The experimental validation and further discussion are respective conducted in Section V and Section VI. Finally, Conclusions are drawn in Section VII.

\section{Fractional-Order Lithium-ion Battery Model}

The fractional-order calculus was first introduced by Leibniz in 1695 to deal with non-integer integrals and derivatives \cite{BWTSC2017}. The fractional-order operator is mathematically defined by ${\sideset{_a}{_t^\mu}{\mathop{\frak{D}}}}$ with $\mu$ being the fractional-order, and $a$, $t$ separately being the lower and upper time bounds \cite{XHTVT2018}. For brevity, ${\sideset{_a}{_t^\mu}{\mathop{\frak{D}}}}$ is simplified as $\frak{D}^\mu$ in the sequel. Based on the Gr{\"u}nwald-Letnikov (GL) definition for fractional derivatives \cite{XHTVT2018}, it gives:
\begin{align}
\label{eq1}
\frak{D}^{\mu}f(t) = \mathop{\text{lim}}\limits_{T_s \rightarrow 0} \frac{1}{T_s^\mu} \sum \limits_{j=0}^{[t/T_s]} \left(-1\right)^j \left \langle \mu, j \right \rangle  f(t-j T_s)
\end{align}
where $f(t)$ denotes a general function at time $t$, $T_s$ is the sampling interval, $[t/T_s]$ is the rounding of $t/T_s$ representing the memory length. With $\Gamma(\cdot)$ as the Gamma function, the Newton binomial coefficient $\left \langle \mu, j \right \rangle$ is defined by
\begin{align*}
\left \langle \mu, j \right \rangle =\left\{
\begin{array}{cl}
1,~&j=0\\
\frac{\Gamma(\mu+1)}{\Gamma(j+1) \cdot \Gamma(\mu-j+1)},~&j>0.
\end{array}
\right.
\end{align*}

With a concern of computational cost, the continuous-time infinite-dimensional GL derivative is discretized and truncated with a short and fixed memory length $L$. As a result, \eqref{eq1} can be approximated as
\begin{align}
\label{eq2}
\frak{D}^{\mu}f_k = \frac{1}{T_s^{\mu}}\sum\limits_{j=0}^L (-1)^j \left \langle \mu, j \right \rangle \cdot f_{k-j}
\end{align}
with $f_k$ being the discretized form of $f(t)$.

\begin{figure}
\centering
\includegraphics[width=0.35\textwidth]{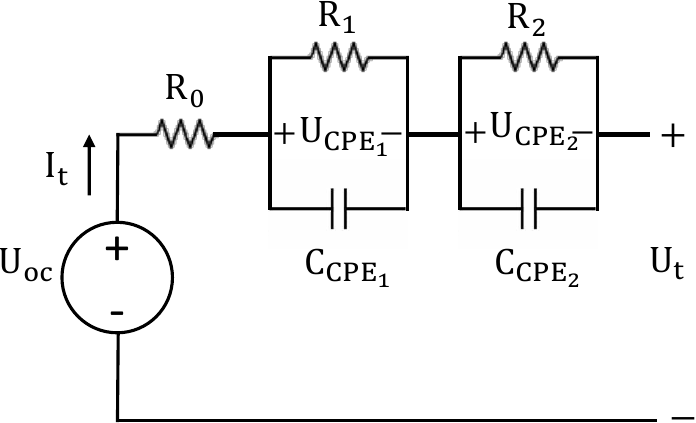}
\caption{FOM with two $RC_{CPE}$ networks}
\label{Fig1}
\end{figure}

The concurrent multiple coupling dynamic is the prominent characteristic of the LIBs during operation. To overcome the shortage of finite RC pair in describing the frequency domain impedance of LIBs, a fractional-order model (FOM) with two constant phase elements (CPEs), shown in Fig. \ref{Fig1} is employed to capture battery behavior. Based on Kirchhoff's law, the dynamic of polarization voltage across two $RC_{CPE}$ networks denoted as $U_{CPE_1}$ and $U_{CPE_2}$ can be given as
\begin{align}
\frak{D}^\alpha U_{CPE_{1,k}} = -\frac{1}{R_1C_{CPE_1}}U_{CPE_{1,k-1}} + \frac{1}{C_{CPE_1}}I_{k-1}\\
\frak{D}^\beta U_{CPE_{2,k}} = -\frac{1}{R_2C_{CPE_2}}U_{CPE_{2,k-1}} + \frac{1}{C_{CPE_2}}I_{k-1}
\end{align}
where $\alpha$ and $\beta$ are the fractional orders of two CPEs, respectively; and $I_{k-1}$ is the load current at time step $k-1$, which is defined to be positive for discharging and negative for charging.

Referring to \eqref{eq2}, the state-space equations can be reformulated as
\begin{align}
\label{eq5}
\nonumber U_{CPE_{1,k}} = &\left(\alpha-\frac{T_s^\alpha}{R_1C_{CPE_1}}\right) U_{CPE_{1,k-1}}+\frac{T_s^\alpha}{C_{CPE_1}}I_{k-1}\\
&-\sum\limits_{j=2}^L \left(-1\right)^j \left \langle \alpha, j \right \rangle \cdot U_{CPE_{1,k-j}}\\
\label{eq6}
\nonumber U_{CPE_{2,k}} = &\left(\beta-\frac{T_s^\beta}{R_2C_{CPE_2}}\right) U_{CPE_{2,k-1}}+\frac{T_s^\beta}{C_{CPE_2}}I_{k-1}\\
&-\sum\limits_{j=2}^L \left(-1\right)^j \left \langle \beta, j \right \rangle \cdot U_{CPE_{2,k-j}}.
\end{align}

The battery SOC can be calculated by the Coulomb Counting method as:
\begin{align}
\label{eq7}
z_k = z_{k-1}-\frac{\eta T_s}{C_n}I_{k-1},
\end{align}
where $z_k$ denotes the battery SOC at time step $k$, $C_n$ denotes the nominal battery capacity and $\eta$ denotes the coulombic efficiency, which is often chosen as 1.
The terminal voltage $U_t$ can be given as
\begin{align}
\label{eq8}
U_{t,k} = U_{oc,k}(z_k) - U_{CPE_{1,k}} -U_{CPE_{2,k}} - R_0 I_k.
\end{align}
The open-circuit voltage $U_{oc}$ is nonlinear, but monotonically increases in terms of SOC. In this work, we apply a six-order polynomial to fit the OCV-SOC curve as
\begin{align}
\nonumber U_{oc}(z) = &a_0 z^0 +a_1 z^1 + a_2 z^2 +  a_3 z^3 \\
&+a_4 z^4 + a_5 z^5 + a_6 z^6
\end{align}
with $a_i, i = 1, \cdots, 6$ denoting the polynomial coefficients. Through the incremental OCV test, these parameters of the selected battery can be identified and summarized in Table \ref{tab1}.

\begin{table}[!t]
\begin{center}
\caption{OCV-SOC polynomial coefficients}
\label{tab1}
\begin{tabular}{c  c c c }
\hline
\hline
Para. & Value &
Para. & Value\\
\hline
$a_0$ & 3.2009 & $a_4$ & -30.7914\\
$a_1$ & 3.9360 & $a_5$ & 5.5057\\
$a_2$ & -16.8149 & $a_6$ & 3.3186\\
$a_3$ & 35.8125&\\
\hline
\hline
\end{tabular}
\end{center}
\end{table}

Considering \eqref{eq5}, \eqref{eq6}, \eqref{eq7} and \eqref{eq8}, a cell-level state-space equations can be formulated as
\begin{align}
\label{eq10}
x_{k+1} =& A x_k + B u_k - \sum \limits_{j=2}^{L+1}(-1)^j \Phi_j x_{k-j+1} + w_k\\
\label{eq11}
y_k =& h(z_k)+Cx_k + D u_k  + v_k
\end{align}
where $h(z_k)=U_{oc,k}(z_k)$, $x_k  = [U_{CPE_{1,k}}, U_{CPE_{2,k}}, z_k]^T$, $y_k = U_{t,k}$, $u_k = I_k$, $w_k$ and $v_k$ are the process noise and measurement noise, respectively. The parameter matrices $A$, $B$ $C$, $D$ and $\Phi$ are defined as follows:
\begin{align}
\nonumber A =& \text{diag}\left(\alpha-\frac{T_s^\alpha}{R_1C_{CPE_1}}, \beta-\frac{T_s^\beta}{R_2C_{CPE_2}}, 1 \right), \\
\nonumber B =& \left[\frac{T_s^\alpha}{C_{CPE_1}}, \frac{T_s^\beta}{C_{CPE_2}}, -\frac{\eta T_s}{C_n}\right],~C = [-1,-1,0]\\
\nonumber \Phi_j =& \text{diag}\left(\left \langle \alpha, j \right \rangle, \left \langle \beta, j \right \rangle, \left \langle 1, j \right \rangle\right),~D = -R_0.
\end{align}

To meet the voltage requirement, LIBs are connected in series to compose a battery system. Referring to the state-space equations \eqref{eq10} and \eqref{eq11}, a battery system with a series connection of $N$ cells can be modeled as:
\begin{align}
\label{eq12}
\mathbf{{x}_{k+1}} = &\textbf{A} \mathbf{{x}_k} + \textbf{B} \mathbf{{u}_k} - \sum \limits_{j=2}^{L+1} (-1)^j \mathbf{\Phi}_j \mathbf{{x}_{k-j+1}}+\mathbf{{w}_k} \\
\label{eq13}
\mathbf{{y}_{k}} =& \mathbf{h(z_k)} + \textbf{C}\mathbf{{x}_k} + \textbf{D} \mathbf{{u}_k} + \mathbf{{v}_k}
\end{align}
where $\mathbf{h(z_k)}=\left[h^1(z_k^1), \cdots, h^N(z_k^N)\right]^T$, $\mathbf{{u}_k} =\left[u_k^1, \cdots, u_k^N\right]^T$ with $u_k^i=I_k^i$, $\mathbf{y_k} = \left[y_k^1, \cdots, y_k^N \right]^T$ with $y_k^i=U_{t,k}^i$, $\mathbf{{w}_k} =\left[w_k^1,\cdots, w_k^N\right]^T$, $\mathbf{{v}_k} =\left[v_k^1,\cdots, v_k^N\right]^T$, $\mathbf{{x}_{k+1}}=\left[x_{k}^1,\cdots, x_{k}^N\right]^T$ with $x_{k}^1=\left[U^1_{CPE_{1,k}}, U^1_{CPE_{2,k}}, z_k^1\right]^T$. The parameter matrices are listed as below:
\begin{align}
\nonumber \textbf{A} = &\text{diag}\left(A^1, \cdots, A^N\right), \textbf{B} = \text{diag}\left(B^1,\cdots,B^N\right),\\
\nonumber \textbf{C} = &\text{diag}\left(C^1,\cdots,C^N\right), \textbf{D} = \text{diag}\left(D^1,\cdots,D^N\right),\\
\nonumber \mathbf{\Phi}_j =&\text{diag}\left(\Phi_j^1,\cdots, \Phi_j^N\right),
\end{align}
where, $A^i$, $B^i$, $C^i$, $D^i$, $\Phi_j^i$ in a system-level model correspond to the matrices found in the cell-level state-space model.

\section{Formulation of Optimization-Based Battery Balancing Problems}

\begin{figure}[!t]
\centering
\includegraphics[width=0.49\textwidth]{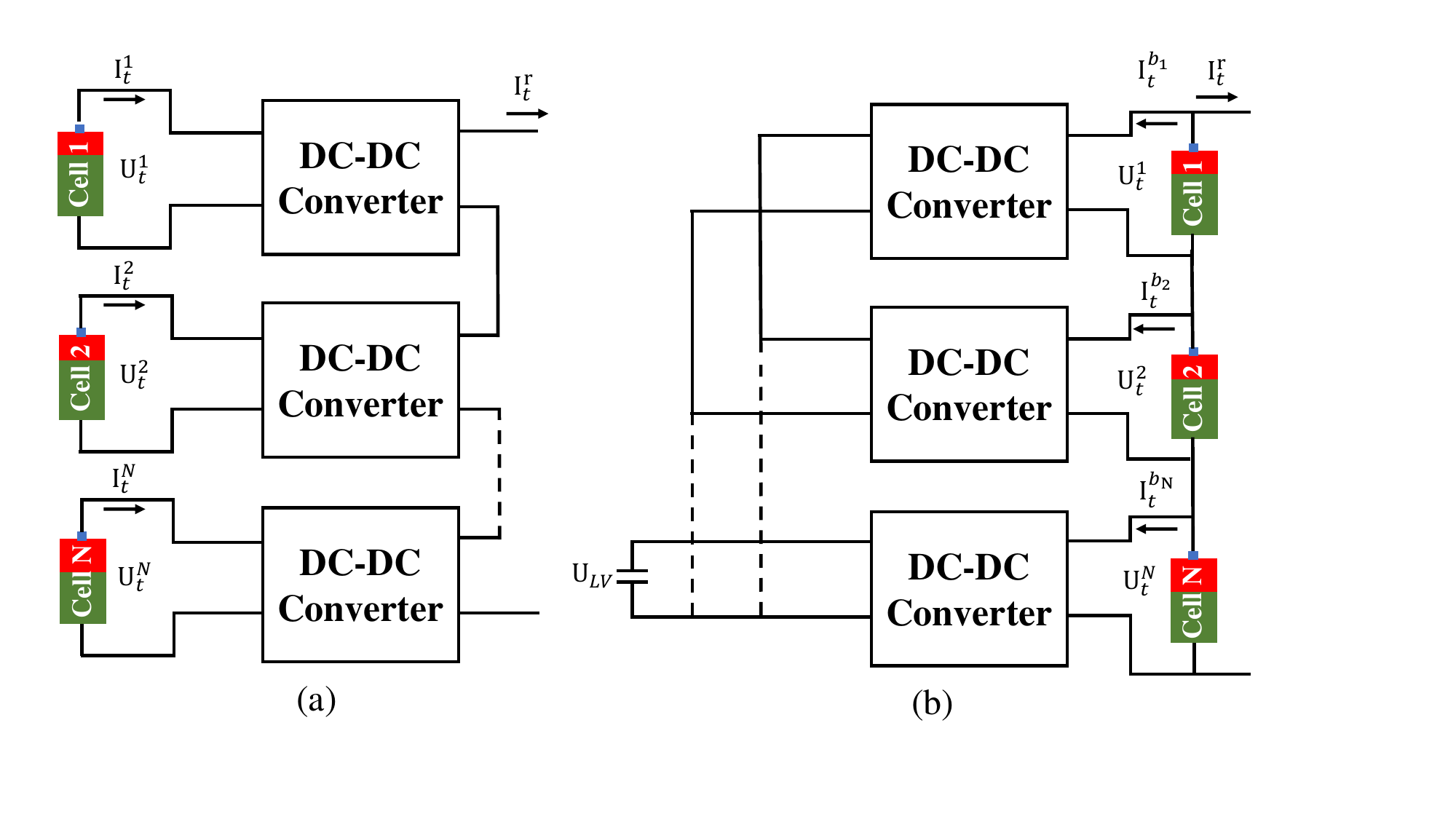}
\caption{(a) Independent cell balancing topology; (b) Differential cell balancing topology}
\label{Fig2}
\end{figure}

As indicated in Fig. \ref{Fig2}, this study examines two distinct cell-level balancing topologies: independent cell balancing and differential cell balancing. The details of the modeling and the current optimization algorithm are elaborated upon in the following sections.

\subsection{Independent Cell Balancing Topology}

Different from the conventional series system where battery cells share a common current $\mathbf{u_k^r}$, the independent cell balancing topology in Fig. \ref{Fig2}(a) realizes the decentralized current control of individual cells via the isolated DC-DC converter. Each battery cell delivers a variable current to its corresponding DC-DC converters according to its estimated SOC; subsequently,  the converters are serially connected to furnish a consistent current to the system, thereby allowing for optimal operation at a cell level.

Based on this balancing topology, we present a control strategy to determine the current distribution $\mathbf{{u}_{k+1}}$ for cell-level balancing to extend the operational time of the battery system. Recalling that the cut-off voltage of a battery system is determined by the lowest cell voltage, an optimization control strategy based on the MPC is devised to actively manipulate the voltages of all the cells and push them away from the cut-off voltage while simultaneously achieving the desired output power and ensuring all the cells to operate within the multi-constraints.

To achieve this goal, the optimal control problem is formulated as
\begin{align}
\label{eq14}
\mathop{\text{min}} \limits_{\mathbf{u_{k+1}}}  &\left(-\mathop{\text{min}} \limits_{i} \left(\hat{y}^i_{k+1}\right)+ \left(\mathbf{\hat{P}_{k+1}}-\mathbf{P_{k+1}}\right)^T\left(\mathbf{\hat{P}_{k+1}}-\mathbf{P_{k+1}}\right)\right) \\
\nonumber \text{s.t.}&\\
 &u_{min} \leq {u}^i_{k+1} \leq u_{max},~i = 1, \cdots, N\\
 &y_{min} \leq {\hat{y}}^i_{k+1} \leq y_{max},~i = 1, \cdots, N
\end{align}
with $\hat{y}_{k+1}^i$ denoting the predicted terminal voltage, $\mathbf{\hat{P}_{k+1}}$ denoting the predicted output power and $\mathbf{{P}_{k+1}}$ denoting the desired output power at time $k+1$.

To reformulate \eqref{eq14}, we introduce a new variable $\epsilon$ to represent the real-time minimum voltage bound. Then the objective function can be rewritten as
\begin{align}
\mathop{\text{min}} \limits_{\mathbf{u_{k+1}},\epsilon}  J(\mathbf{u_k},\epsilon)=-\epsilon + \left(\mathbf{\hat{P}_{k+1}}-\mathbf{P_{k+1}}\right)^T\left(\mathbf{\hat{P}_{k+1}}
-\mathbf{P_{k+1}}\right)\label{eq17}
\end{align}
with an additional constraint
\begin{align}
\epsilon \leq \hat{y}^i_{k+1},~~ i =1,\cdots,N.
\end{align}

Furthermore, the battery voltage prediction is formulated as:
\begin{align}
\label{eq19}
\mathbf{\hat{x}_{k+1}} =& \mathbf{A} \mathbf{\hat{x}_{k}}+ \mathbf{B} \mathbf{u_k}- \sum \limits_{j=2}^{L+1} (-1)^j \mathbf{\Phi}_j \mathbf{\hat{x}}_{k-j+1}\\
\label{eq20}
\mathbf{\hat{y}_{k+1}} =& \mathbf{h}(\mathbf{\hat{z}_k})+ \mathbf{C} \mathbf{\hat{x}_{k+1}} + \mathbf{D} \mathbf{u_{k+1}}.
\end{align}

When predicting output power of battery system $\mathbf{\hat{P}_{k+1}}$, we use the reference current $\mathbf{u_{k+1}^r}$ and it becomes:
\begin{align}
\mathbf{\hat{P}_{k+1}} = \left( \mathbf{h}(\mathbf{\hat{z}_k})+\mathbf{C} \mathbf{\hat{x}_{k+1}} + \mathbf{D} \mathbf{u^r_{k+1}}\right) \mathbf{u_{k+1}}.
\end{align}

For brevity, we denote $ \mathbf{h}(\mathbf{\hat{z}_k})+ \mathbf{C} \mathbf{\hat{x}_{k+1}} + \mathbf{D} \mathbf{u^r_{k+1}}$ as $\mathbf{\hat{y}_{k+1}^r}$.

The cost function \eqref{eq17} can be rewritten as:
\begin{align}
\label{eq22}
 \nonumber \mathop{\text{min}} \limits_{\mathbf{u_{k+1}},\epsilon}&  J(\mathbf{u_k},\epsilon)\\
\nonumber = &-\epsilon+\left(\mathbf{\hat{y}_{k+1}^{r^T}}\mathbf{u_{k+1}}-\mathbf{P_{k+1}}\right)^T\left(\mathbf{\hat{y}_{k+1}^{r^T}}\mathbf{u_{k+1}}-\mathbf{P_{k+1}}\right)\\
\nonumber =&-\epsilon + \frac{1}{2}\mathbf{{u}_{k+1}}^T 2 \mathbf{\hat{y}_{k+1}^{r}} \mathbf{\hat{y}_{k+1}^{r^T}}  \mathbf{{u}_{k+1}}+\mathbf{P_{k+1}}\mathbf{P_{k+1}}\\
\nonumber &-2\mathbf{P_{k+1}}\mathbf{\hat{y}_{k+1}^{r^T}} \mathbf{{u}_{k+1}}\\
= & \frac{1}{2} \begin{bmatrix}
\mathbf{u_{k+1}}\\
\epsilon
\end{bmatrix}^T
2 \underbrace{\text{diag}\left(\mathbf{\hat{y}_{k+1}^{r^T}}\mathbf{\hat{y}_{k+1}^{r}}, 0\right)}_{W_{mpc}}
\begin{bmatrix}
\mathbf{u_{k+1}}\\
\epsilon
\end{bmatrix}\\
\nonumber &+\underbrace{\left[-2\mathbf{P_{k+1}}\mathbf{\hat{y}_{k+1}^{r^T}}, -1\right]}_{V_{mpc}}
\begin{bmatrix}
\mathbf{u_{k+1}}\\
\epsilon
\end{bmatrix}
\end{align}

Ignoring the variation of $\mathbf{U_{oc}}$ at time $k$ and $k+1$, the matrix inequality for voltage constraints can be given as \eqref{eq23}.

\begin{figure*}[!h]
\begin{align}
\label{eq23}
\begin{bmatrix}
\mathbf{D} & \mathbf{0}\\
-\mathbf{D} & \mathbf{0}\\
-\mathbf{D} & \mathbf{I}
\end{bmatrix}
\begin{bmatrix}
\mathbf{u_{k+1}}\\
\epsilon
\end{bmatrix} \leq
\begin{bmatrix}
\mathbf{y_{max}}- \mathbf{h}(\mathbf{\hat{z}_k})-\mathbf{C}\left(\mathbf{A}\mathbf{\hat{x}_k}+\mathbf{B}\mathbf{u_k}- \sum \limits_{j=2}^{L+1} (-1)^j \mathbf{\Phi}_j \mathbf{\hat{x}}_{k-j+1}\right)\\
-\mathbf{y_{min}}+ \mathbf{h}(\mathbf{\hat{z}_k})+\mathbf{C}\left(\mathbf{A}\mathbf{\hat{x}_k}+\mathbf{B}\mathbf{u_k}- \sum \limits_{j=2}^{L+1} (-1)^j \mathbf{\Phi}_j \mathbf{\hat{x}}_{k-j+1}\right)\\
 \mathbf{h}(\mathbf{\hat{z}_k})+\mathbf{C}\left(\mathbf{A}\mathbf{\hat{x}_k}+\mathbf{B}\mathbf{u_k}- \sum \limits_{j=2}^{L+1} (-1)^j \mathbf{\Phi}_j \mathbf{\hat{x}}_{k-j+1}\right)
\end{bmatrix}
\end{align}
\end{figure*}

Similarly, the current constraint can be constructed as:
\begin{align}
\label{eq24}
\begin{bmatrix}
\mathbf{I} & \mathbf{0}\\
-\mathbf{I} & \mathbf{0}\\
\end{bmatrix}
\begin{bmatrix}
\mathbf{u_{k+1}}\\
\epsilon
\end{bmatrix}\leq \begin{bmatrix}
\mathbf{u_{max}}\\
-\mathbf{u_{min}}\\
\end{bmatrix}.
\end{align}

Besides, the SOC constraints can be formulated as \eqref{eq25}.

\begin{figure*}[!h]
\begin{align}
\label{eq25}
\begin{bmatrix}
\mathbf{H} \mathbf{B} & \mathbf{0}\\
\mathbf{H} \mathbf{B} & \mathbf{0}
\end{bmatrix}
\begin{bmatrix}
\mathbf{u_{k+1}}\\
\epsilon
\end{bmatrix} \leq
\begin{bmatrix}
\mathbf{z_{max}}-\mathbf{H}\mathbf{A} \left(\mathbf{A} \mathbf{\hat{x}_k} +\mathbf{B} \mathbf{u_k}- \sum \limits_{j=2}^{L+1} (-1)^j \mathbf{\Phi}_j \mathbf{\hat{x}}_{k-j+1}\right)- \sum \limits_{j=1}^{L} (-1)^j \mathbf{\Phi}_j \mathbf{\hat{x}}_{k-j+1}\\
-\mathbf{z_{min}}+\mathbf{H}\mathbf{A} \left(\mathbf{A} \mathbf{\hat{x}_k} +\mathbf{B} \mathbf{u_k}- \sum \limits_{j=2}^{L+1} (-1)^j \mathbf{\Phi}_j \mathbf{\hat{x}}_{k-j+1}\right)- \sum \limits_{j=1}^{L} (-1)^j \mathbf{\Phi}_j \mathbf{\hat{x}}_{k-j+1}
\end{bmatrix},
\end{align}
\end{figure*}

In \eqref{eq25}, $\mathbf{H} = \text{diag}\left(\left[e_1,\cdots,e_1\right]\right)$ with $e_1=\left[0,0,1\right]$, $\mathbf{z_{max}}$ and $\mathbf{z_{min}}$ are the battery upper and lower cut-off SOCs, respectively.

Given \eqref{eq23}, \eqref{eq24} and \eqref{eq25}, the multi-constraint matrix inequality can be constructed as \eqref{eq26}.
\begin{figure*}
\begin{align}
\label{eq26}
\underbrace{\begin{bmatrix}
\mathbf{I} & \mathbf{0}\\
-\mathbf{I} & \mathbf{0}\\
\mathbf{H} \mathbf{B} & \mathbf{0}\\
\mathbf{H} \mathbf{B} & \mathbf{0}\\
\mathbf{D} & \mathbf{0}\\
-\mathbf{D} & \mathbf{0}\\
-\mathbf{D} & \mathbf{I}
\end{bmatrix}}_{A_{mpc}}
\begin{bmatrix}
\mathbf{u_{k+1}}\\
\epsilon
\end{bmatrix} \leq
\underbrace{\begin{bmatrix}
\mathbf{u_{max}}\\
-\mathbf{u_{min}}\\
\mathbf{z_{max}}-\mathbf{H}\mathbf{A} \left(\mathbf{A} \mathbf{\hat{x}_k} +\mathbf{B} \mathbf{u_k}- \sum \limits_{j=2}^{L+1} (-1)^j \mathbf{\Phi}_j \mathbf{\hat{x}}_{k-j+1}\right)- \sum \limits_{j=1}^{L} (-1)^j \mathbf{\Phi}_j \mathbf{\hat{x}}_{k-j+1}\\
-\mathbf{z_{min}}+\mathbf{H}\mathbf{A} \left(\mathbf{A} \mathbf{\hat{x}_k} +\mathbf{B} \mathbf{u_k}- \sum \limits_{j=2}^{L+1} (-1)^j \mathbf{\Phi}_j \mathbf{\hat{x}}_{k-j+1}\right)- \sum \limits_{j=1}^{L} (-1)^j \mathbf{\Phi}_j \mathbf{\hat{x}}_{k-j+1}\\
\mathbf{y_{max}}- \mathbf{h}(\mathbf{\hat{z}_k})-\mathbf{C}\left(\mathbf{A}\mathbf{\hat{x}_k}+\mathbf{B}\mathbf{u_k}- \sum \limits_{j=2}^{L+1} (-1)^j \mathbf{\Phi}_j \mathbf{\hat{x}}_{k-j+1}\right)\\
-\mathbf{y_{min}}+ \mathbf{h}(\mathbf{\hat{z}_k})+\mathbf{C}\left(\mathbf{A}\mathbf{\hat{x}_k}+\mathbf{B}\mathbf{u_k}- \sum \limits_{j=2}^{L+1} (-1)^j \mathbf{\Phi}_j \mathbf{\hat{x}}_{k-j+1}\right)\\
 \mathbf{h}(\mathbf{\hat{z}_k})+\mathbf{C}\left(\mathbf{A}\mathbf{\hat{x}_k}+\mathbf{B}\mathbf{u_k}- \sum \limits_{j=2}^{L+1} (-1)^j \mathbf{\Phi}_j \mathbf{\hat{x}}_{k-j+1}\right)
\end{bmatrix}}_{B_{mpc}}
\end{align}
\end{figure*}

\subsection{Differential Cell Balancing Topology}

The differential cell balancing topology, shown in Fig. \ref{Fig2}(b), connects the DC-DC bypass converters in parallel with battery cells at the output port and shares the input port with the LV bus voltage $U_{LV}$ as the common reference point for the energy exchange. The bypass converters operate in a distributed fashion by discharging or charging different currents based on the SOC of each cell, thereby enabling the series-connected cells to supply power to the system at different rates.

Hence, the battery system prediction in the state space form of \eqref{eq19} and \eqref{eq20} can be reconstructed as
\begin{align}
\label{eq27}
\mathbf{\hat{x}_{k+1}} =& \mathbf{A} \mathbf{\hat{x}_{k}}+ \mathbf{B} (\mathbf{u^r_k}+\mathbf{u^b_k})- \sum \limits_{j=2}^{L+1} (-1)^j \mathbf{\Phi}_j \mathbf{\hat{x}}_{k-j+1}\\
\label{eq28}
\mathbf{\hat{y}_{k+1}} =& \mathbf{h}(\mathbf{\hat{z}_k})+ \mathbf{C} \mathbf{\hat{x}_{k+1}} + \mathbf{D} (\mathbf{u^r_{k+1}}+\mathbf{u^b_k})
\end{align}
where $\mathbf{u^b_k} =\left[u^{b_1}_k, \cdots, u^{b_N}_k\right]^T$ with $u^{b_i}_k$ being the balance current flowing from the bypass converter $i$ at time step $k$.

The optimal control problem is then formulated as
\begin{align}
\label{eq29}
\mathop{\text{min}} \limits_{\mathbf{u_{k+1}}}  &\left(-\mathop{\text{min}} \limits_{i} \left(\hat{y}^i_{k+1}\right)+ \left(\mathbf{\hat{P}_{k+1}}-\mathbf{P_{k+1}}\right)^T
\left(\mathbf{\hat{P}_{k+1}}-\mathbf{P_{k+1}}\right)\right)\\
\nonumber \text{s.t.}&\\
 &u_{min} \leq {u}^i_{k+1} \leq u_{max},~i = 1, \cdots, N\label{eq29-1}\\
 &y_{min} \leq {\hat{y}}^i_{k+1} \leq y_{max},~i = 1, \cdots, N\label{eq29-2}\\
 &\sum \limits_{i=1}^N u^{b_i}_k = 0.\label{eq29-3}
\end{align}

Following a similar procedure, the matrices for the current distribution can be obtained as \eqref{eq33} and \eqref{eq34}.

\begin{figure*}
\begin{align}
\label{eq33}
&W_{mpc} = 2 \text{diag} \left(\mathbf{\hat{y}_{k+1}^{r^T}}\mathbf{\hat{y}_{k+1}^{r}}, 0\right), V_{mpc} = 2 \left[\mathbf{\hat{y}_{k+1}^{r^T}}\mathbf{{u}_{k+1}^{r}}-\mathbf{P_{k+1}},-1\right]\\
\label{eq34}
 & A_{mpc} = \begin{bmatrix}
\mathbf{I} & \mathbf{0}\\
-\mathbf{I} & \mathbf{0}\\
\mathbf{H} \mathbf{B} & \mathbf{0}\\
\mathbf{H} \mathbf{B} & \mathbf{0}\\
\mathbf{D} & \mathbf{0}\\
-\mathbf{D} & \mathbf{0}\\
-\mathbf{D} & \mathbf{I}
\end{bmatrix}, B_{mpc} = \begin{bmatrix}
\mathbf{u_{max}}\\
-\mathbf{u_{min}}\\
\mathbf{z_{max}}-\mathbf{H}\mathbf{A} \left(\mathbf{A} \mathbf{\hat{x}_k} +\mathbf{B} (\mathbf{u^r_k}+\mathbf{u^b_k})- \sum \limits_{j=2}^{L+1} (-1)^j \mathbf{\Phi}_j \mathbf{\hat{x}}_{k-j+1}\right)- \sum \limits_{j=1}^{L} (-1)^j \mathbf{\Phi}_j \mathbf{\hat{x}}_{k-j+1}\\
-\mathbf{z_{min}}+\mathbf{H}\mathbf{A} \left(\mathbf{A} \mathbf{\hat{x}_k} +\mathbf{B} (\mathbf{u^r_k}+\mathbf{u^b_k})- \sum \limits_{j=2}^{L+1} (-1)^j \mathbf{\Phi}_j \mathbf{\hat{x}}_{k-j+1}\right)- \sum \limits_{j=1}^{L} (-1)^j \mathbf{\Phi}_j \mathbf{\hat{x}}_{k-j+1}\\
\mathbf{y_{max}}- \mathbf{h}(\mathbf{\hat{z}_k})-\mathbf{C}\left(\mathbf{A}\mathbf{\hat{x}_k}+\mathbf{B}(\mathbf{u^r_k}+\mathbf{u^b_k})- \sum \limits_{j=2}^{L+1} (-1)^j \mathbf{\Phi}_j \mathbf{\hat{x}}_{k-j+1}\right)+\mathbf{D}\mathbf{u^r_{k+1}}\\
-\mathbf{y_{min}}+ \mathbf{h}(\mathbf{\hat{z}_k})+\mathbf{C}\left(\mathbf{A}\mathbf{\hat{x}_k}+\mathbf{B}(\mathbf{u^r_k}+\mathbf{u^b_k})- \sum \limits_{j=2}^{L+1} (-1)^j \mathbf{\Phi}_j \mathbf{\hat{x}}_{k-j+1}\right)-\mathbf{D}\mathbf{u^r_{k+1}}\\
 \mathbf{h}(\mathbf{\hat{z}_k})+\mathbf{C}\left(\mathbf{A}\mathbf{\hat{x}_k}+\mathbf{B}(\mathbf{u^r_k}+\mathbf{u^b_k})- \sum \limits_{j=2}^{L+1} (-1)^j \mathbf{\Phi}_j \mathbf{\hat{x}}_{k-j+1}\right)+\mathbf{D}\mathbf{u^r_{k+1}}
\end{bmatrix}
\end{align}
\end{figure*}

With the known matrices $W_{mpc}$, $V_{mpc}$, $A_{mpc}$ and $B_{mpc}$, the multi-constraint optimization problem can be cast into quadratic programming problem that is efficiently solvable by $\textit{quadprog}$ in MATLAB, enabling the determination of the optimal current distribution and the minimum cell terminal voltage.

\section{Online Current Distribution Algorithms}

Based on the derived battery system state space equations \eqref{eq12} and \eqref{eq13}, the desired online current distribution algorithms for PISO topology, namely Algorithm \ref{algo1}, and for SIPO topology, namely Algorithm \ref{algo2}, are designed in this section, respectively.

\begin{algorithm}[!t]
    \caption{Current distribution for the independent cell balancing topology}
\label{algo1}
    \begin{algorithmic}[1]
	\ENSURE{branch current $\mathbf{u}$}
        \REQUIRE{reference current $\mathbf{u^r}$, Power demand $\mathbf{P}$}

\STATE \textit{initialize} simulation time $T_{max}$, number of series-connected batteries $N$, polynomial coefficients for $U_{oc}$, battery parameter matrices ($\mathbf{A}$, $\mathbf{B}$, $\mathbf{C}$, $\mathbf{D}$), initial estimate $\mathbf{\hat{x}_0}=\left[\hat{x}^1_0, \cdots, \hat{x}^N_0\right]^T$, initial branch currents $\mathbf{u_0}$;

\FOR{$k = 0:1:T_{max}$}
\STATE\COMMENT{$k$ is the sampling index}

\STATE \textit{collect} $\mathbf{P_{k+1}}$;
\STATE \textit{collect} $\mathbf{u^r_{k+1}}$;
\STATE \textit{extract} $\mathbf{\hat{z}_k}$ from $\mathbf{\hat{x}_k}$;

$\backslash$\% \textbf{Optimization-based current determination} \%$\backslash$
\STATE \textit{update} $\mathbf{\hat{x}_{k+1}}$ based on \eqref{eq19};
\STATE \textit{extract} $\mathbf{\hat{z}_{k+1}}$ from $\mathbf{\hat{x}_{k+1}}$;
\STATE \textit{update} $\mathbf{h(\hat{z}_{{k+1}})}$;
\STATE \textit{calculate} $\mathbf{\hat{y}_{k+1}^{r}}$ according to $\mathbf{u^r_{k+1}}$;
\STATE \textit{update} matrices $W_{mpc}$, $V_{mpc}$ in \eqref{eq22}, $A_{mpc}$, $B_{mpc}$ in \eqref{eq26};
\STATE \textit{solve} the constrained optimization problem \eqref{eq14};
\STATE \textit{extract} $\mathbf{u_{k+1}}$;

$\backslash$\% \textbf{EKF-based state estimation} \%$\backslash$\\
\STATE \textit{measure} $\mathbf{y_{k+1}}$ under $\mathbf{u_{k+1}}$;
\STATE \textit{calculate} $\mathbf{\hat{y}_{k+1}}$ from $\mathbf{h(\hat{z}_{k+1})} + \textbf{C}\mathbf{\hat{x}_{k+1}} + \textbf{D} \mathbf{{u}_{k+1}}$;

\STATE\COMMENT{time update}

\STATE \textit{update} $\textit{d}U_{oc, k+1}/\textit{d}t$;
\STATE \textit{update} gain matrix $\mathbf{K}$ and covariance matrices;
\STATE \textit{update} $\mathbf{\hat{x}_{k+1}}=\mathbf{\hat{x}_{k+1}}+ \mathbf{K}(\mathbf{y_{k+1}-\hat{y}_{k+1}}$);
\STATE\COMMENT{measurement update}
\ENDFOR
    \end{algorithmic}
\end{algorithm}

\begin{algorithm}[!t]
    \caption{Current distribution for the differential cell balancing topology}
\label{algo2}
    \begin{algorithmic}[1]
	\ENSURE{balance current $\mathbf{u^b}$}
        \REQUIRE{reference current $\mathbf{u^r}$, Power demand $\mathbf{P}$}

\STATE \textit{initialize} simulation time $T_{max}$, number of series-connected batteries $N$, polynomial coefficients for $U_{oc}$, battery parameter matrices ($\mathbf{A}$, $\mathbf{B}$, $\mathbf{C}$, $\mathbf{D}$), initial estimate $\mathbf{\hat{x}_0}=\left[\hat{x}^1_0, \cdots, \hat{x}^N_0\right]^T$, initial balance current $\mathbf{u^b_0}$;

\FOR{$k = 0:1:T_{max}$}
\STATE\COMMENT{$k$ is the sampling index}

\STATE \textit{collect} $\mathbf{P_{k+1}}$;
\STATE \textit{collect} $\mathbf{u^r_{k+1}}$;
\STATE \textit{extract} $\mathbf{\hat{z}_k}$ from $\mathbf{\hat{x}_k}$;

$\backslash$\% \textbf{Optimization-based current determination} \%$\backslash$
\STATE \textit{update} $\mathbf{\hat{x}_{k+1}}$ based on \eqref{eq27};
\STATE \textit{extract} $\mathbf{\hat{z}_{k+1}}$ from $\mathbf{\hat{x}_{k+1}}$;
\STATE \textit{update} $\mathbf{h(\hat{z}_{{k+1}})}$;
\STATE \textit{calculate} $\mathbf{\hat{y}_{k+1}^{r}}$ according to $\mathbf{u^b_{k+1}}+\mathbf{u^r_{k+1}}$;
\STATE \textit{update} matrices $W_{mpc}$, $V_{mpc}$ in \eqref{eq33}, $A_{mpc}$, $B_{mpc}$ in \eqref{eq34};
\STATE \textit{solve} the constrained optimization problem \eqref{eq29};

$\backslash$\% \textbf{EKF-based state estimation} \%$\backslash$\\
\STATE \textit{measure} $\mathbf{y_{k+1}}$ under $\mathbf{u^b_{k+1}}+\mathbf{u^r_{k+1}}$;
\STATE \textit{calculate} $\mathbf{\hat{y}_{k+1}}$ from $\mathbf{h(\hat{z}_{k+1})} + \textbf{C}\mathbf{\hat{x}_{k+1}} + \textbf{D} (\mathbf{{u}^r_{k+1}}+\mathbf{{u}^b_{k+1}})$;
\STATE\COMMENT{time update}
\STATE \textit{update} $\textit{d}U_{oc, k+1}/\textit{d}t$;
\STATE \textit{update} gain matrix $\mathbf{K}$ and covariance matrices;
\STATE \textit{update} $\mathbf{\hat{x}_{k+1}}=\mathbf{\hat{x}_{k+1}}+ \mathbf{K}(\mathbf{y_{k+1}-\hat{y}_{k+1}})$;
\STATE\COMMENT{measurement update}
\ENDFOR
    \end{algorithmic}
\end{algorithm}

Both Algorithms 1 and 2 consist of two parts: (1) MPC-based current distribution optimization; and (2) extended Kalman filter (EKF)-based state estimation. The former aims to use MPC to regulate cell currents while satisfying the power demand and pushing the minimal cell terminal voltage away from the cut-off voltage. The latter aims to employ EKF to iteratively estimate battery SOC based on the derived current and minimize the estimation error coming from erroneous initialization according to the difference between the predicted and measured cell terminal voltage.
It is worth noting that in EKF-based state estimation the prediction of cell terminal voltage at time $k+1$, namely time update, necessitates the current distribution. Therefore, we first employ MPC to obtain the current distribution at $k+1$ time step and then use it for subsequent EKF-based state estimation. The readers can refer to \cite{GPEKF} for the details about the design of the EKF-based state estimation.

By comparing algorithms between the independent cell balancing topology and the differential cell balancing topology, the main distinction lies in the controlling target.
For independent cell balancing, it regulates the individual cell output power by controling the current for each cell, denoted as $\mathbf{u}$. In contrast, for differential cell balancing, the cell-level output power management is achieved through the adjustment of balance current $\mathbf{u^b}$ from DC-DC bypass converters.

\section{Experimental Validation}

\begin{figure}[!h]
\centering
\includegraphics[width=0.42\textwidth]{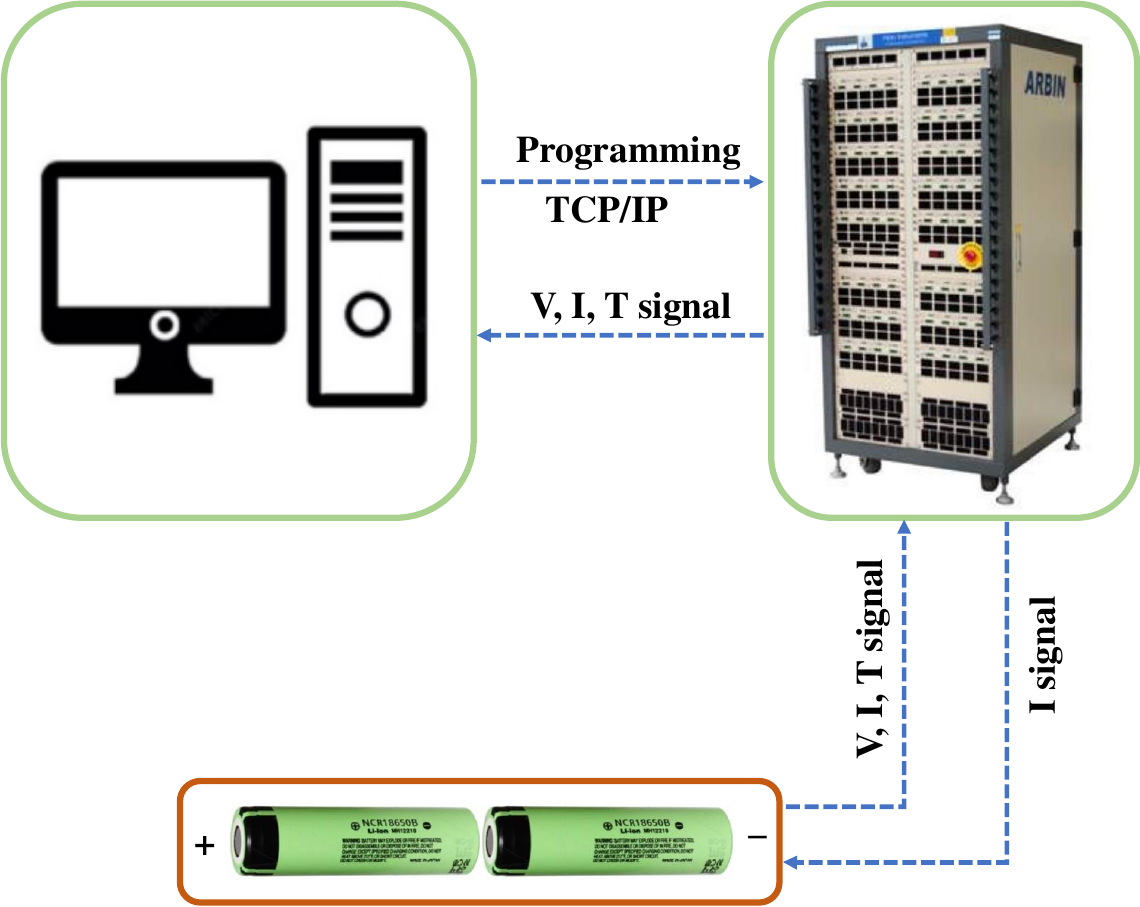}
\caption{Experimental setup}
\label{Fig3}
\end{figure}

Experiments have been carried out on a commercial LIB, NCR18650B and the detailed battery specification is listed in Table. \ref{tab2}. The configuration of a battery test bench is established in Fig. \ref{Fig3}, where an Arbin BT2000 system is utilized to provide current profiles based on EV driving cycles and a host computer is employed to set up the testing program and collect data generated during the experiments.

\begin{table}[!t]
\begin{center}
\caption{Specifications of NCR18650B Battery}
\label{tab2}
\begin{tabular}{c | c }
\hline
\hline
Items &Specification\\
\hline
Nominal capacity & 3.2~Ah\\
Nominal voltage & 3.6~V\\
Maximum discharge current & 6.4~A\\
Operating voltage range & 3~V - 4.2~V\\
\hline
\hline
\end{tabular}
\end{center}
\end{table}

\begin{figure}
\centering
\includegraphics[width=0.49\textwidth]{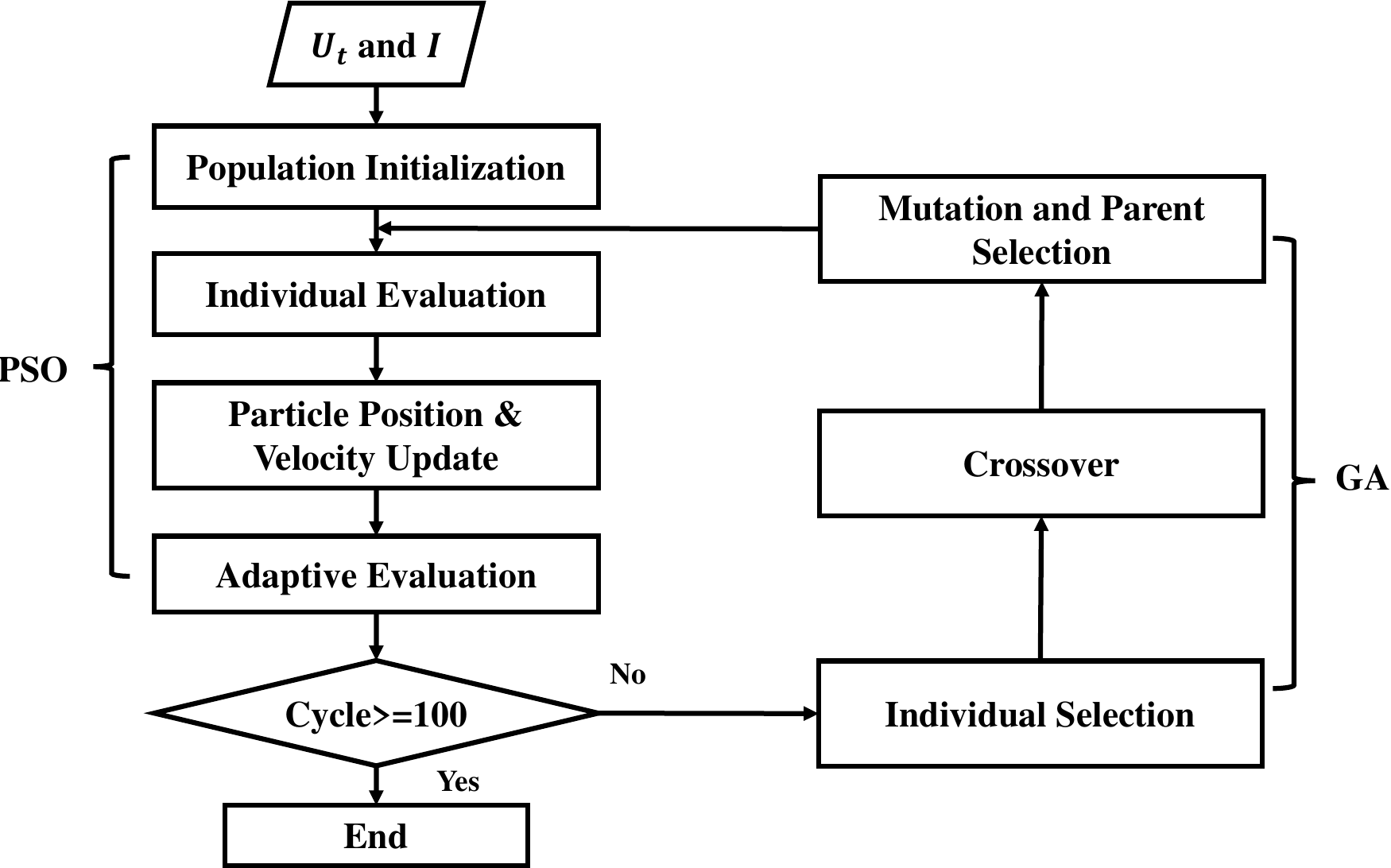}
\caption{Flow chart of the PSO-GA method}
\label{Fig4}
\end{figure}

To identify the FOM parameters
\begin{align}
\nonumber \Theta = \left[R_{0}, R_1, R_2, C_{CPE_1}, C_{CPE_2}, \alpha, \beta \right]^T,
\end{align}
a particle swarm optimization-genetic algorithm (PSO-GA) method is employed \cite{CFPSO}. It replaces the particle position/velocity updates with the selection and mutation for optimal solution derivation so that the candidate solutions can rapidly converge to the global optimal solution with less probability of being trapped into local minima. The operating procedure of the PSO-GA method is shown in Fig. \ref{Fig4}. The PSO-GA method aims to minimize the sum of the squared errors between the measured and predicted battery terminal voltages as
\begin{align}
\text{min} f\left(\hat{\Theta}\right) = \sum \limits_{k=1}^{M}\left[U_{t,k}-\hat{U}_{t,k}(x_k, u_k, \hat{\Theta})\right].
\end{align}

\begin{figure}[!t]
\centering
\includegraphics[width=0.45\textwidth]{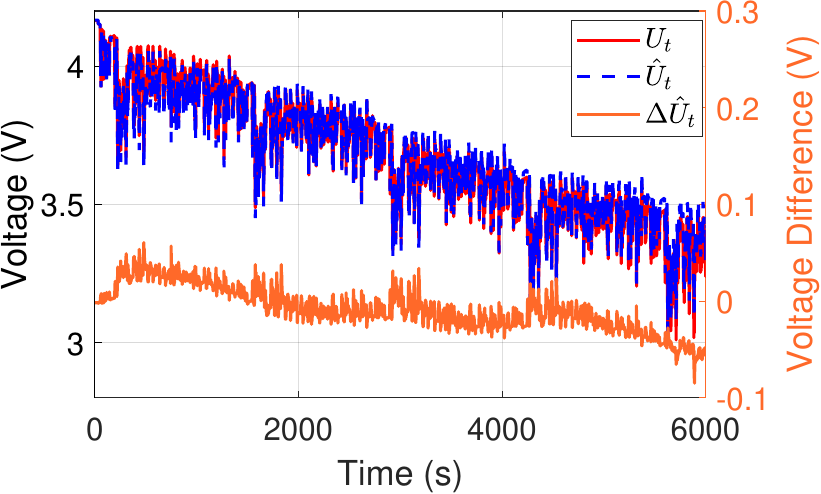}
\caption{Model validation for PSO-GA based battery parameter identification}
\label{Fig5}
\end{figure}

Based on the experimental results under the current profile based on UDDS, the voltage fitting curve is shown in Fig. \ref{Fig5}. According to the voltage difference between the estimation and the measurement (indicated by orange solid line), it is clear that the second-order FOM model can accurately follow the measurement trajectories, thereby confirming the precision of the modeling.

Afterwards, we take a two-series battery system discharging under the current profile based on UDDS as an example for illustration. A four-step verification procedure is then carried out to demonstrate the efficacy of the designed method:

\begin{itemize}
\item \textbf{Step 1:} Discharge the two-series battery  system under the current profile based on UDDS and then record the current and voltage;
\item \textbf{Step 2:} Calculate the power consumption for two cells and take the calculation as power demand;
\item \textbf{Step 3:} Implement the design algorithm \ref{algo1} or \ref{algo2} to determine the branch current $\mathbf{u}$ or balance current $\mathbf{u^b}$ to extending battery system operational time;
\item \textbf{Step 4:} Update the new current profiles and discharge the same battery  system under two different topologies for comparison.
\end{itemize}

\section{Results and Discussion}
\begin{figure}
\centering
\includegraphics[width=0.48\textwidth]{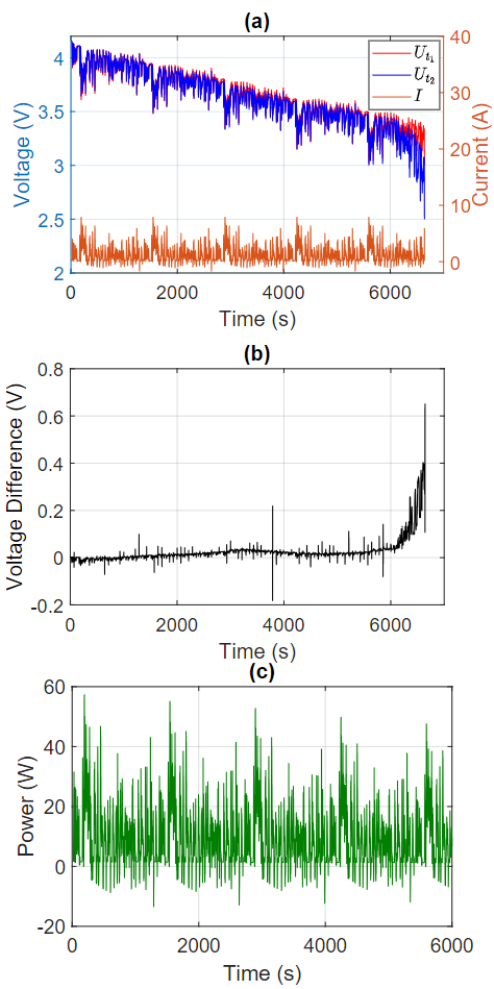}
\caption{(a) Voltage and current for reference sharing current topology; (b) Voltage error; (c) Power demand}
\label{Fig6}
\end{figure}

In this section, we aim to showcase the effectiveness of the proposed cell balancing algorithm in both independent cell balancing and differential cell balancing topologies. The battery parameters for the two cells are listed in Table \ref{tab3}. The reference terminal voltage and power under the common sharing current are demonstrated in Fig. \ref{Fig6}(a)-(c).
\begin{table}[!t]
\begin{center}
\caption {Parameter Identification Results of a two-series connected battery system}
\label{tab3}
\begin{tabular}{c | c | c }
\hline
\hline
 & Cell~1 & Cell~2\\
\hline
$R_{0}$ & 0.0545 $\Omega$& 0.0567 $\Omega$\\
\hline
$R_1$ & 0.4567 $\Omega$ & 0.4314 $\Omega$\\
\hline
$C_{CPE_1}$ & 4950 $F$ & 4999.7 $F$\\
\hline
$R_2$ & 0.4959 $\Omega$ & 0.0137 $\Omega$\\
\hline
$C_{CPE_2}$ & 270.6 $F$ & 802.1$F$\\
\hline
$\alpha$ & 0.3110 & 0.9103\\
\hline
$\beta$ &  0.0548& 0.061\\
\hline
\hline
\end{tabular}
\end{center}
\end{table}
Due to the inconsistency of battery parameters, under the same working profiles, the voltage response of cells can exhibit variations. As shown in Fig. \ref{Fig6}(a), the cutoff for battery operation is determined by the minimal cell terminal voltage within the system. Consequently, even if Cell 1 (indicated in red solid line) has sufficient energy, it is the voltage of Cell 2 (indicated in blue solid line) that determines the cutoff for the series battery system.

\begin{figure}[!t]
\flushleft
~~\includegraphics[width=0.43\textwidth]{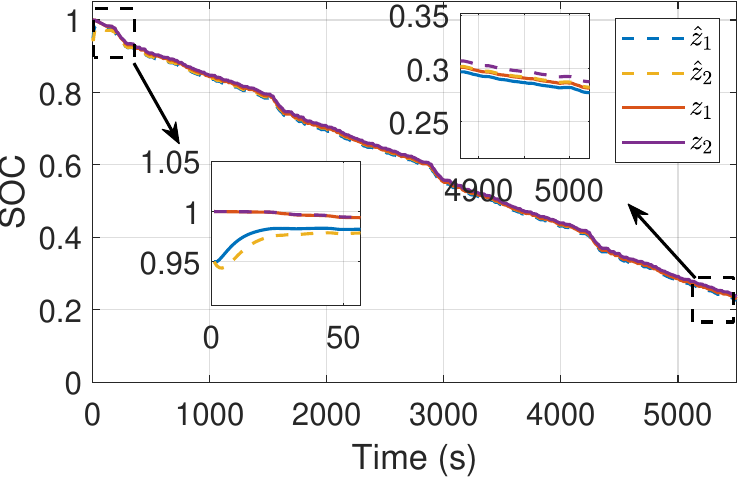}
\caption{Battery SOC estimation and reference}
\label{Fig7}
\end{figure}

Discussions on the performance of Algorithms 1 and 2 will be elaborated from the two perspectives: EKF-based state estimation and optimization-based current determination. As for EKF-based state estimation, the initial covariance matrices of the process noise and measurement noise are configured to be
\begin{align}
\nonumber P = \begin{bmatrix}
0.1 & 0 & 0\\
0 & 0.1& 0\\
0.1 & 0.1& 0.1
\end{bmatrix},
Q = \begin{bmatrix}
10^{-2} & 0 & 0 \\
0 & 10^{-2} & 0\\
0 & 0& 10^{-9}
\end{bmatrix}
\end{align}
and the initial SOC is set to 95$\%$, where the true SOC is 100 $\%$). The experimental results of online SOC estimation are shown in Fig. \ref{Fig7}. The reference SOC calculated by coulomb counting is indicated in a solid line while the EKF-based SOC estimation is indicated in a dashed line. Despite the presence of initial SOC estimation errors, it is evident that the estimation can converge to the true SOC in a fast manner with a high level of accuracy. Such accurate SOC estimation establishes a solid foundation to predict cell terminal voltage and current distribution.

\begin{figure}[!t]
\centering
\includegraphics[width=0.45\textwidth]{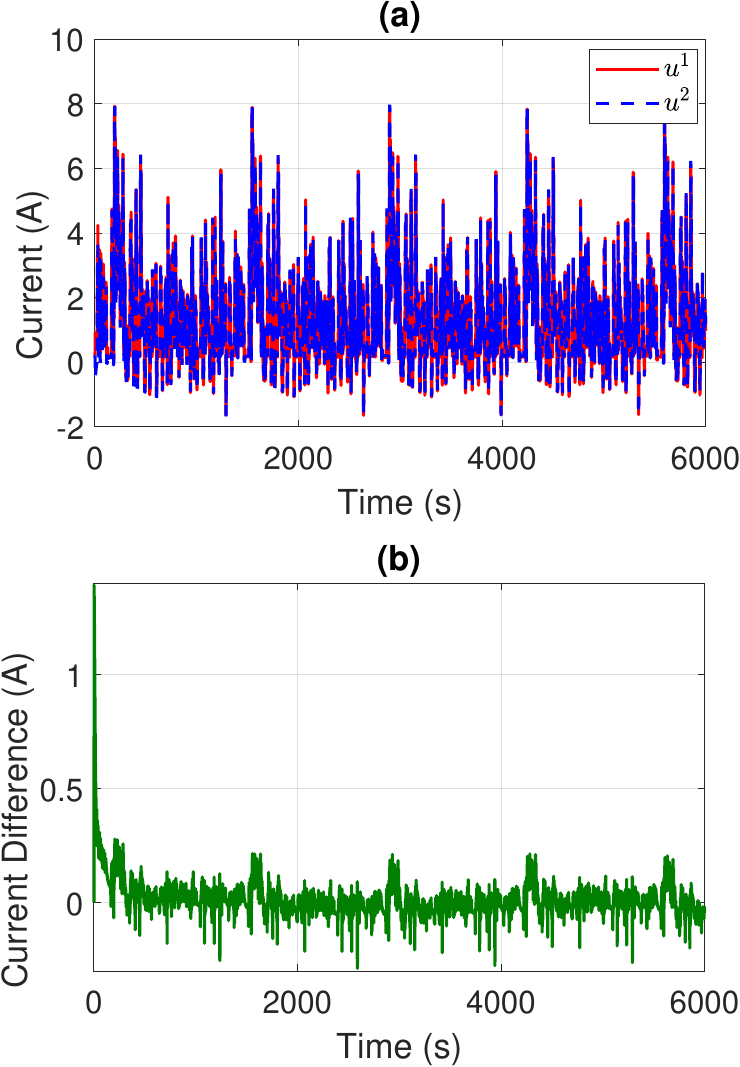}
\caption{(a) Current distribution for Cell 1 and Cell 2 under PISO topology; (b) Current difference between Cell 1 and Cell 2 under PISO topology}
\label{Fig8}
\end{figure}

\begin{figure}[!h]
\centering
\includegraphics[width=0.45\textwidth]{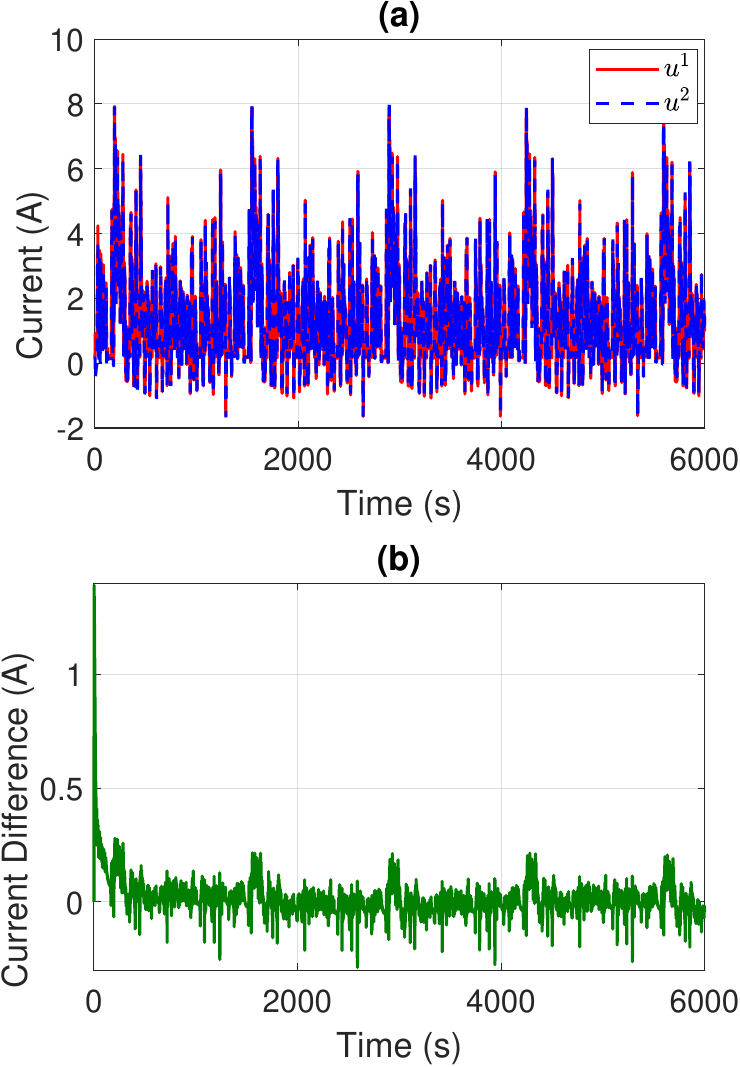}
\caption{(a) Current distribution for Cell 1 and Cell 2 under SIPO topology; (b) Current difference between Cell 1 and Cell 2 under SIPO topology}
\label{Fig9}
\end{figure}
\begin{figure}[!h]
\centering
\includegraphics[width=0.45\textwidth]{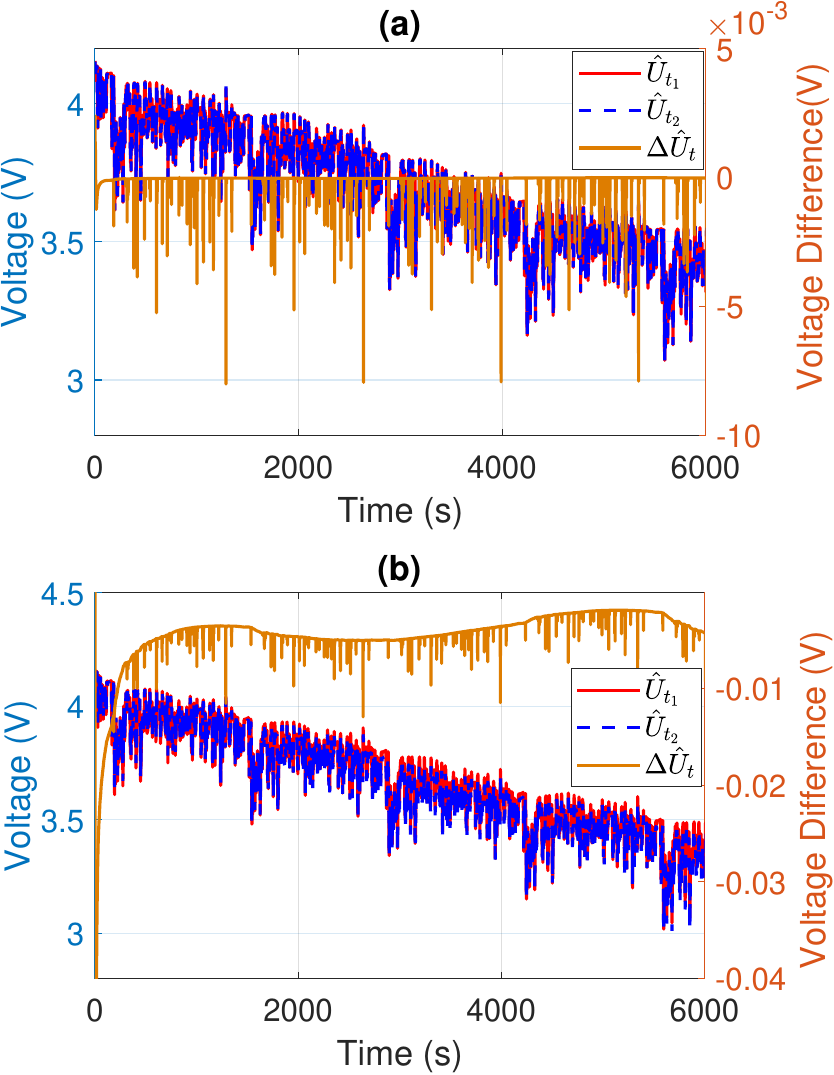}
\caption{(a) Terminal voltages for Cell 1 and Cell 2 under PISO topology; (b) Terminal voltages for Cell 1 and Cell 2 under SIPO topology}
\label{Fig10}
\end{figure}
\begin{figure}[!h]
\centering
\includegraphics[width=0.45\textwidth]{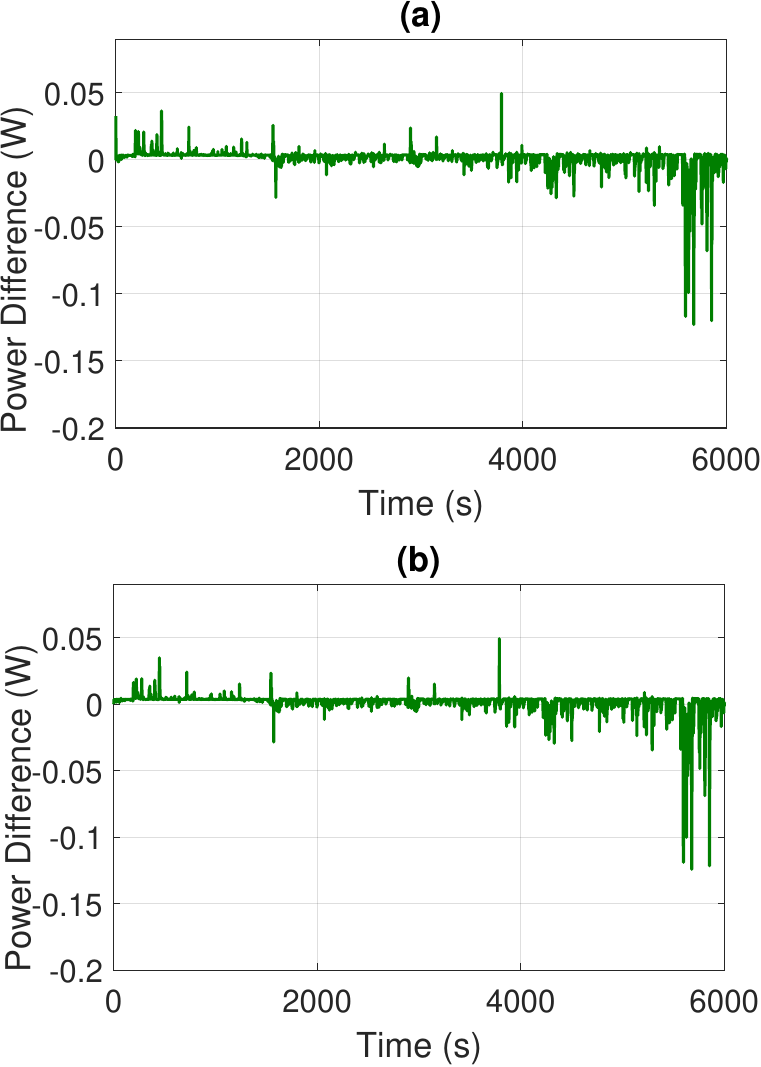}
\caption{(a) Power difference under PISO topology; (b) Power difference under SIPO topology}
\label{Fig11}
\end{figure}

Regarding optimization-based current determination, with the same EKF parameters and power demand, the updated current profiles and the current difference for two different balancing topologies are respectively illustrated in Figs. \ref{Fig8} and \ref{Fig9}. Owing to the presence of inconsistent battery parameters, the current flowing through Cell 1 characterized by a low internal resistance is larger than that of Cell 2 characterized by a high internal resistance. The resultant terminal voltages under two cases are presented in Fig. \ref{Fig10}. By comparing with Fig. \ref{Fig6}(b), it can be clearly seen that  the MPC based optimization provides a noticeable reduction in voltage disparity between Cell 1 and Cell 2. Additionally, the minimal terminal voltage of the cells can maintain above the cut-off voltage, thereby extending the operational time for the series-connected battery system.
To examine the operational time extension enabled by the proposed method, an analysis is conducted on the SOC of cells at the end of discharging. Before implementing the current optimization algorithm, the SOCs for Cell 1 and Cell 2 were respectively recorded at 0$\%$ and 5.5$\%$, while after balancing, the SOC for both cells was equalized to 3.2$\%$, which substantiates the efficacy of the balancing algorithm.
Furthermore, Fig. \ref{Fig11} demonstrates the power difference between the expected and actual power output. The root-mean-square error for the independent cell balancing is 0.0101 $W$ while that for the differential cell balancing is 0.0102 $W$, which confirms that the proposed method can meet the power demand while extending the operational time of the series-connected system.

\section{Conclusions}
In this study, we introduce an active balancing strategy by leveraging MPC to optimize the minimum cell voltage and thereby prolong the operational time of the battery system. The method integrates fractional order battery system modeling with PSO-GA parameter identification, effectively capturing the characteristics of battery systems. By addressing the constraints related to voltage, current and power as well as solving the optimal control problems, the distribution of cell currents in a battery system can be determined. The experimental validations are implemented under two distinct cell-level balancing topologies. Experimental results demonstrate that the proposed balancing method can achieve an approximate 3.2$\%$ service time extension.

\end{document}